\documentclass[10pt]{bmc_article}

\usepackage{cite} 
\usepackage{hyperref}  
\usepackage{ifthen}  
\usepackage{multicol}   
\usepackage[utf8]{inputenc} 
\usepackage{color}

\usepackage{graphicx}
\usepackage{caption}
\usepackage{subfigure}
\newboolean{publ}

\newenvironment{bmcformat}{\baselineskip20pt\sloppy\setboolean{publ}{false}}{\baselineskip20pt\sloppy}

\begin{document}
\begin{bmcformat}


\title{Agalma: an automated phylogenomics workflow}
 

\author{
Casey W. Dunn\correspondingauthor$^1$%
\email{Casey W. Dunn\correspondingauthor - casey\_dunn@brown.edu},
Mark Howison$^{1,2}$%
\email{Mark Howison - mhowison@brown.edu},
and 
Felipe Zapata$^1$%
\email{Felipe Zapata - felipe\_zapata@brown.edu}
}


\address{%
    \iid(1)Department of Ecology and Evolutionary Biology, Brown University, Providence, RI, USA\\
    \iid(2)Center for Computation and Visualization, Brown University, Providence, RI, USA
}%

\maketitle


\begin{abstract}
\textbf{Background:}
In the past decade, transcriptome data have become an important component of many phylogenetic studies. They are a cost-effective source of protein-coding gene sequences, and have helped projects grow from a few genes to hundreds or thousands of genes. Phylogenetic studies now regularly include genes from newly sequenced transcriptomes, as well as publicly available transcriptomes and genomes. Implementing such a phylogenomic study, however, is computationally intensive, requires the coordinated use of many complex software tools, and includes multiple steps for which no published tools exist. Phylogenomic studies have therefore been manual or semiautomated. In addition to taking considerable user time, this makes phylogenomic analyses difficult to reproduce, compare, and extend. In addition, methodological improvements made in the context of one study often cannot be easily applied and evaluated in the context of other studies.

\textbf{Results:}
We present Agalma, an automated tool that conducts phylogenomic analyses. The user provides raw Illumina transcriptome data, and Agalma produces annotated assemblies, aligned gene sequence matrices, a preliminary phylogeny, and detailed diagnostics that allow the investigator to make extensive assessments of intermediate analysis steps and the final results.  Sequences from other sources, such as externally assembled genomes and transcriptomes, can also be incorporated in the analyses. Agalma is built on the BioLite bioinformatics framework, which tracks provenance, profiles processor and memory use, records diagnostics, manages metadata, and enables rich HTML reports for all stages of the analysis. Agalma includes a small test data set and a built-in test analysis of these data. In addition to describing Agalma, we here present a sample analysis of a larger seven-taxon data set. Agalma is available for download at \url{https://bitbucket.org/caseywdunn/agalma}.

\textbf{Conclusions:}
Agalma allows complex phylogenomic analyses to be implemented and described unambiguously as a series of high-level commands. This will enable phylogenomic studies to be readily reproduced, modified, and extended. Agalma also facilitates methods development by providing a complete modular workflow, bundled with test data, that will allow further optimization of each step in the context of a full phylogenomic analysis. 

\end{abstract}

\ifthenelse{\boolean{publ}}{\begin{multicols}{2}}{}


\section*{Background}

Transcriptome data are fast becoming an important and cost effective component of phylogenetic studies \cite{Dunn:2008ky, Hejnol2009, Smith:2011de, Philippe:2009hh, Delsuc:2006cw}. The rapid fall in sequencing prices has contributed to the growing number of phylogenetic studies that integrate data from genomes and transcriptomes, often referred to as  ``phylogenomic'' analyses. There is wide recognition that adding data from a larger number of genes is necessary to address many open phylogenetic questions, though of course additional gene sequences alone will not be sufficient to resolve them all \cite{Sanderson:2008jz, Philippe:2006cx, Edgecombe:2011cw}.

The primary impediments to wider adoption and further improvement of phylogenomic methods are the complexity of the analyses and the lack of integrated tools to conduct them. Each phylogenomic study requires many steps, the vast majority of which concern matrix construction rather than phylogenetic analysis itself. These steps include raw data filtering, assembly, identification of ribosomal RNA, selection of transcript splice variants, translation, identification of homologous sequences, identification of orthologous sequences, sequence alignment, phylogenetic analysis, and summary of results. Implementing a phylogenomic analysis is not just a matter of executing available tools for each of these steps. Among other challenges, results must be summarized across multiple steps, detailed records must be kept of all analysis steps, data files often need to be reformatted between analyses, and computational load must be balanced according to the available resources. 

Because phylogenomic studies are complex and have been manual or semi-automated, they are difficult to implement and explicitly describe, and require extensive technical effort to reproduce. These problems can make it difficult to evaluate results, integrate data across studies, expand analyses, or test the impact of alternative analysis approaches. In addition, manual analyses often include many subjective decisions that may impact the final results. 

Some higher-level pipelines have addressed subsets of phylogenomic analyses. These tools include PartiGene \cite{Parkinson:2004uh}, a pipeline to aid in the assembly and annotation of Sanger transcriptome data collected across a diversity of species, and SCaFoS \cite{Roure:2007hi}, a semi-automated tool for the curation of super matrices from previously assembled transcriptomes. No existing tool, however, can execute a full phylogenomic analysis of modern sequence data.

We addressed these needs by developing Agalma, an automated phylogenomics workflow. Using Agalma, an investigator can conduct complete phylogenomic analyses, from raw sequence reads to preliminary phylogenetic trees, with a small number of high-level commands. The results are accompanied by detailed reports that integrate diagnostic information across data sets and analysis steps. In a first pass with Agalma, the investigator conducts the analysis under default settings. The investigator then consults the reports to consider how best to optimize the analyses, and easily re-executes them with updated settings to produce new matrices and preliminary trees. The investigator can then analyze the optimized matrices with external phylogenetic inference tools not already included within Agalma to explore other factors, such as model sensitivity. 

\section*{Implementation}

We built Agalma with BioLite~\cite{Howison2012}, a generic bioinformatics pipeline
framework. BioLite profiles memory and CPU use, tracks provenance of all data and analyses, and
centralizes diagnostic reporting. Agalma is a modular workflow comprised of helper scripts and a series of pipelines. Each
pipeline is made up of stages that call internal analysis functions (many implemented with the help of Biopython~\cite{cock_biopython:_2009}) and
wrappers from the BioLite Python module.  The wrappers invoke
command-line tools, which include external bioinformatics tools such as the
Bowtie2 aligner \cite{Langmead:2012jh} and Trinity assembler \cite{Grabherr:2011jb}, as well as several C++ tools from BioLite.

The first step for analyzing each data set, whether it consists of raw reads to be assembled or of previously assembled gene predictions, is to catalog the data. This creates a database entry that includes sample metadata and the paths to the data files.  Agalma has built-in support for transcriptome assembly of pair-end Illumina data only.   When analyzing public data, raw reads and associated metadata can be imported directly from the NCBI Sequence Read Archives (SRA) using the command \texttt{sra import}.  This command downloads the reads for a given SRA accession number (experiment, study, sample, or run), converts them into FASTQ format, and populates  the catalog with the corresponding data paths and metadata.

There are several distinct tasks subsequent to cataloging the data--sequence assembly, loading the genes into the database, and phylogenetic analysis. These tasks are described in detail in the \texttt{README} and \texttt{TUTORIAL} files provided with Agalma, and are briefly summarized below.

\subsection*{Assembly}

The pipeline \texttt{transcriptome} runs an assembly from read filtering through assembly and preliminary annotation. In a typical analysis, \texttt{transcriptome} would be run once for each species for which raw Illumina transcriptome data are available.
The \texttt{transcriptome} pipeline executes the following sub-pipelines, which can also be run individually:

\begin{itemize}
\item \texttt{sanitize} filters raw paired-end Illumina data and randomizes the order of reads (maintaining read order between paired files) to
facilitate subsequent subsetting. Reads are discarded if they do not meet a
specified mean quality threshold, if they contain Illumina adapter sequences,
or if the base composition is highly skewed (if any base represents either $<5\%$ or $>60\%$
of the sequence).  This pipeline also generates FastQC~\cite{andrews_fastqc} summaries of read quality.

\item \texttt{insert\_size} uses subassemblies and mapping to estimate the mean
and variance of the insert size (i.e., the length of the fragment between the sequencing adapters). This information provides important feedback on the success of sample preparation, and is also used in some downstream analysis steps.

\item \texttt{remove\_rrna} removes most reads that are derived from ribosomal RNA (rRNA). Removing rRNA in advance of assembly can reduce both the number of chimeric transcripts and the time required for assembly. This pipeline first assembles multiple subsets of reads. A range of subset sizes is used since the optimal 
number of reads for assembling a particular rRNA transcript depends upon multiple factors, 
including the fraction of reads that originate from rRNA and the uniformity of coverage across the rRNA transcripts (which can vary greatly, depending on how denatured the samples were prior to fragmentation). 
rRNA transcripts are then identified by blast comparison of these subassemblies to an included dataset of known rRNA sequences.
The entire set of reads is then compared to the rRNA transcripts that are identified form the subassemblies, and any
reads that map to them are removed. A plot of the distribution of reads across exemplar rRNA transcripts is shown to help evaluate rRNA assembly success. The top hit 
in the NCBI \texttt{nt} database is also provided as an independent check on sample identity and to help spot potential contaminants. The fraction of reads that derive from rRNA is also reported to aid in improving library preparations.

\item \texttt{assemble} filters the reads at a higher quality threshold and assembles them. Assemblies can be conducted under multiple protocols (such as multiple assemblers, or the same assembler under different settings). This pipeline can also assemble multiple subsets of different numbers of reads, which provides perspective on how sequencing effort impacts assembly results. The default assembler is Trinity \cite{Grabherr2011}. The wrapper we have included in Agalma for running Trinity makes two improvements over the wrapper script that comes with Trinity. First, we have added a filter in between the Chrysalis and Butterfly stages to remove components that are smaller than the minimum transcript length parameter passed to Butterfly, since running Butterfly on these components will not yield a transcript. For the five assemblies in our test data set, this reduces the number of Butterfly commands from roughly 100,000 to 60,000. Second, we have replaced the ParaFly utility that is used for concurrent execution of the Butterfly commands with the GNU \texttt{parallel} tool \cite{Tange2011} because it has better parallel efficiency. ParaFly executes the commands concurrently, but in blocks, so that the time to execute a block is the runtime of the slowest individual command. The runtimes can vary greatly because of variance in transcript length and complexity. In contrast, \texttt{parallel} load balances the commands across the available processors.

\item \texttt{postassemble} uses \texttt{blastn} against the rRNA reference sequences to identify rRNA transcripts (these could include low abundance transcripts, such as parasite contaminants, that were not removed as reads by \texttt{remove\_rrna}) and screens against the NCBI UniVec database to identify vector
contaminants (such as protein expression vector contaminants in the sample preparation enzymes, which we have encountered in multiple samples). It selects an exemplar transcript (i.e., splice variant) for each gene based on an assembly confidence score. It maps the reads to the exemplar transcripts to build a coverage map that helps assess the distribution of sequencing effort across genes.
Finally, it uses \texttt{blastx} to compare the transcripts against the NCBI SwissProt database to establish which are similar to previously known proteins.

\end{itemize}

Following these steps, the investigator can inspect the assembled data directly or load them into the Agalma database to prepare them for phylogenetic analysis.

\subsection*{Load genes into the local Agalma database}

Subsequent phylogenetic analyses require that all gene sequences to be considered are loaded into the local Agalma database. The 
\texttt{load} command takes care of this process. In a typical analysis, \texttt{load} is executed once for each dataset that has been assembled by the \texttt{transcriptome} pipeline described above, and once for each set of gene predictions from external sources (e.g., externally assembled 454 transcriptome data or gene predictions from genome sequencing projects).

\subsection*{Phylogenetic analysis}

Once  assemblies for multiple species are loaded into the local Agalma database, the user carries out a phylogenomic analysis by consecutively executing the following pipelines:

\begin{itemize}

\item \texttt{homologize} allows the user to specify which datasets to include in a particular phylogenetic analysis. It then uses an all-by-all \texttt{tblastx} search to build a graph with edges representing hits above a stringent threshold, and breaks the graph into connected components corresponding to clusters of homologous sequences with the Markov Clustering Algorithm (MCL) tool \cite{dongen_using_2012}. 

\item \texttt{multalign} applies sampling and length filters to each cluster of homologous sequences, and uses another all-by-all \texttt{tblastx} within each cluster to trim sequence ends that have no similarity to other sequences in the cluster (these could include, for example, chimeric regions). The sequences of each cluster are aligned using MACSE~\cite{Ranwez:2011gk}, a translation-aware multiple-sequence aligner that accounts for frameshifts and stop codons. Multiple sequence alignments are then cleaned with GBLOCKS~\cite{Talavera:2007gi}. Optionally, the alignments can be concatenated together to form a supermatrix.

\item \texttt{genetree} uses RAxML~\cite{Stamatakis:2006de} to build a maximum likelihood phylogenetic tree for each cluster of homologous sequences. Gene trees can be filtered according to mean bootstrap support, which eliminates genes that have little phylogenetic signal \cite{Salichos:2014bx} and reduces overall computational burden. This filter can be applied prior to running \texttt{treeprune} (described below), which has the added advantage of restricting ortholog selection to only well-supported gene trees. All options available in RAxML can be passed as optional arguments. If the input is a supermatrix consisting of concatenated orthologs, it builds a species tree. 

\item \texttt{treeprune} identifies orthologs according to the topology of gene phylogenies, using a new implementation of the method introduced in a previous phylogenomic study \cite{Hejnol:2009bi}. It uses DendroPy~\cite{sukumaran_dendropy:_2010} to prune each gene tree into maximally inclusive subtrees with no more than one sequence per taxon. Each of these subtrees is considered as a set of putative orthologs. The pruned subtrees are re-entered as clusters into Agalma's database.

\end{itemize}

After \texttt{treeprune}, the user can build a supermatrix and a preliminary maximum likelihood species tree with RAxML. These steps, which include rerunning \texttt{multalign} and \texttt{genetree} on the orthologs, are detailed in the Agalma \texttt{TUTORIAL} file. The user can then  proceed with more extensive phylogenetic analyses of the supermatrix using external phylogenetic inference software of their choice (only RAxML is included with Agalma at this time). As the alignments for each gene are also provided, the investigator can also apply promising new approaches that simultaneously infer gene trees and species trees \cite{Boussau:2013ba}.

\subsection*{Test data and analyses}

A small set of test data is provided with Agalma. It consists of 25,000 100bp Illumina read pairs for the siphonophore \emph{Hippopodius hippopus}, a subset of 72-74 gene sequences assembled for each of five siphonophores, and a subset of 40 gene predictions from the \emph{Nematostella vectensis} genome assembly. These data were chosen because they run relatively quickly and enable testing of most commonly used features.

These test data serve several purposes. They allow a user to validate that Agalma is working correctly, and users are strongly encouraged to run this test with the command \texttt{agalma test} right after installation. The test data also serve as the foundation for the example analysis described in the \texttt{TUTORIAL} file. For the developer, the \texttt{agalma test} command serves as a regression test to check if changes break existing features. We routinely run this test in the course of adding and refactoring code. The test data also serve as a minimal case study for developing new features without needing to first download and curate data.

\section*{Results and Discussion}

In addition to the small test data sets included with Agalma, here we present an example analysis of larger data sets from seven species.
Though most phylogenetic analyses would include more taxa than this simple example case, the size of the dataset for each species is typical for contemporary phylogenomic analyses. This seven-taxon data set consists of new raw Illumina reads for five species of siphonophores, \emph{Abylopsis tetragona}, \emph{Agalma elegans}, \emph{Nanomia bijuga}. \emph{Physalia physalis}, \emph{Craseoa sp.}, and gene predictions for two outgroup taxa, \emph{Nematostella vectensis} \cite{Putnam:2007iw} and \emph{Hydra magnipapillata} \cite{Chapman:2010ik}, produced by previous genome projects. mRNA for the five siphonophore samples was isolated with Dynabeads mRNA DIRECT (Life Technologies) and prepared for sequencing with the TruSeq RNA Sample Prep Kit (Illumina). The sample preparation was modified by including a size selection step (agarose gel cut) prior to PCR amplification. Analyses were conducted with Agalma version 0.3.3.

We deposited the new data in a public archive (NCBI Sequence Read Archive, BioProject PRJNA205486) prior to running the final analysis. A git repository of the scripts we used to execute the example analyses is available at \url{https://bitbucket.org/caseywdunn/dunnhowisonzapata2013}. These scripts download the data from the public archives, execute the analyses, and generate the analysis reports. All of the figures presented here are taken from the analysis reports generated by Agalma. This illustrates how a fully reproducible and open phylogenomic analysis can be implemented and communicated with Agalma. These scripts can be used as they are to repeat the analyses. They could also be modified to try alternative analysis strategies on these same data, or they could be adapted to run similar analyses on different data.

\subsection*{Assembly}

The tabular assembly report (\texttt{index.html} in Additional File 1) summarizes assembly statistics across samples, and links to more detailed assembly reports for each sample. For the example analysis, this summary indicates, among other things, that the fraction of rRNA in each library ranged from $0.4$\% to $27.2$\% and the insert sizes were on average $266$ bp long. The detailed assembly reports have extensive diagnostics that pertain to sample quality and the success of library preparation. As an example, Figure 1 shows several of the plots from the detailed assembly report for \emph{Agalma elegans}, the siphonophore after which our tool is named. The distribution of sequencing effort across genes (Figure 1a) and the size distribution of transcripts (Figure 1b) are typical for \emph{de novo} Illumina transcriptome assemblies.

\subsection*{Phylogenetic analyses}

The Agalma phylogeny report includes a plot of the number of genes considered at each stage of the analysis. In the example analysis, the step that removed the most genes was cluster refinement in \texttt{multalign} (Figure 2).  This reduction is largely due to the elimination of clusters that failed the taxon sampling criteria, and reflects uncertainty regarding the homology of some sequences and sparse sampling of some homologs.  The next major reduction in the number of genes occurred in \texttt{treeprune}. These reductions are due to both uncertainty regarding orthology and poor sampling of some ortholog groups. The preliminary species tree for the example analysis (Figure 3) is congruent with previous analyses of siphonophore relationships \cite{Dunn:2005dy}.

\subsection*{Resource Utilization}

Phylogenomic analyses are computationally intensive. Detailed information about resource utilization helps investigators plan resources for projects and balance computational load more efficiently. It is also critical for the optimization of the analyses, and can help guide design decisions. 
For each analysis, Agalma produces a resource utilization plot that displays the time and maximum memory used by external executables (Figure 4). The peak memory use, and the longest-running step, in the `transcriptome` pipeline was \texttt{assemble}.

\section*{Conclusions}

A distinction is sometimes drawn between manual approaches that enable close user inspection of data and results, and automated approaches that isolate the user from their results. This is a false dichotomy--automating analyses and examining the results closely are not mutually exclusive. Automated analyses with detailed diagnostics provide the best of both worlds. The user has a very detailed perspective on their analysis, and the added efficiencies of automation leave the investigator with far more time to assess these results. Automation also allow improvements made in the context of one study to be applied to other studies much more effectively.

For a study to be fully reproducible, both the data and the analysis must be described explicitly and unambiguously. The best description of an analysis is the code that was used to execute the analysis. By automating phylogenomic analyses from data download through matrix construction and preliminary phylogenetic trees, Agalma enables fully reproducible phylogenomic studies. This will allow reviewers and readers to reproduce an entire analysis exactly as it was run by the authors, without needing to re-curate the same dataset or rewrite the analysis code.

There are alternative approaches to many of the steps in a phylogenomic analysis presented here. 
There are, for example, multiple tools that identify orthologs according to different methods and criteria \cite{Chen:2007ft,Altenhoff:2013ep}.
Agalma is a general framework and can be expanded to include these additional methods, and directly compare them in the context of a complete workflow that is consistent in all other regards. 
\bigskip

\section*{Availability and requirements}
\textbf{Project name:} Agalma\\
\textbf{Project home page:} \url{https://bitbucket.org/caseywdunn/agalma}\\
\textbf{Operating system(s):} Linux, Mac OS X\\
\textbf{Programming language:} Python\\
\textbf{Other requirements:} BioLite\\
\textbf{License:} GNU

\section*{Competing interests}
The authors declare that they have no competing interests.

\section*{Author's contributions}
CWD and MH conceived of the original design of Agalma, and together with FZ made most design decisions. 
MH, CWD, and FZ implemented and tested the software and wrote the manuscript. MH did most of the coding. 
MH and FZ conducted the computational analyses.
All authors read and approved the final manuscript.

\section*{Acknowledgements}
  \ifthenelse{\boolean{publ}}{\small}{}
Steve Haddock provided several of the specimens, and Freya Goetz prepared all material for sequencing.
Thanks to Deb Goodwin and the students aboard \emph{SSV Robert C. Seamans} for collecting the \emph{Physalia physalis} specimens, and to Erik Zettler and the Sea Education Association for facilitating this opportunity. Thanks also to Christian Sardet for assisting with collecting the \emph{Hippopodius hippopus}. Freya Goetz prepared the two siphonophore samples for sequencing. 
Stephen Smith and Nicholas Sinnott-Armstrong made important contributions to early planning of Agalma. Stefan Siebert assisted with curation of the test dataset and provided suggestions on features. 
Richard Nguyen provided HTML and CSS styling for the reports.
Thanks to Steve Haddock, Warren Francis,  Lingsheng Dong, and other early-adopters for feedback on previous versions of Agalma.
Rebecca Helm provided helpful comments on an earlier version of this manuscript. 
This research was conducted using computational resources and services at the
Center for Computation and Visualization, Brown University.
Agalma was developed with the support of the US National Science Foundation
(awards 0844596, 0844596, 1004057, and 1256695).
 

\newpage
{\ifthenelse{\boolean{publ}}{\footnotesize}{\small}
 \bibliographystyle{bmc_article}  
  \bibliography{agalma} }     


\ifthenelse{\boolean{publ}}{\end{multicols}}{}



\section*{Figures}

\includegraphics[width=\columnwidth]{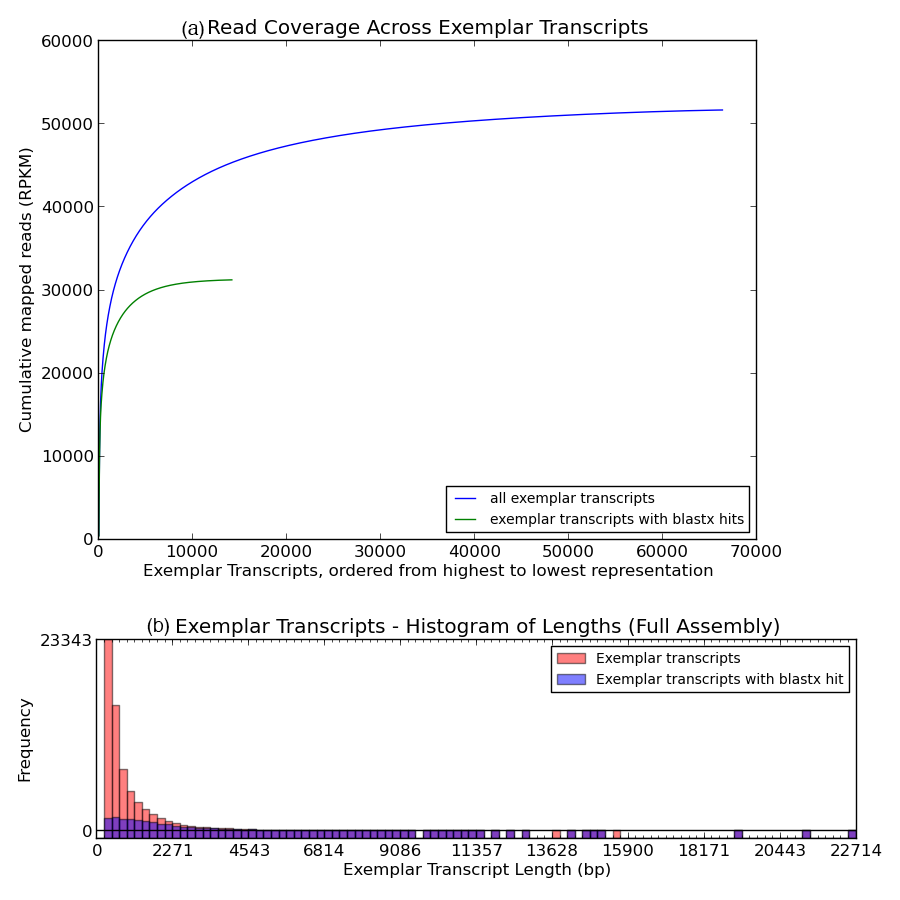}
\subsection*{Figure 1 - Transcriptome assembly statistics for \emph{Agalma elegans}}
(a) The distribution of sequencing reads across genes, sorted from the most frequently sequenced genes to the least frequently sequenced genes. (b) The size distribution of the assembled genes, for all genes and for only those genes that have a blastx hit to a protein in the swissprot database. See Additional File 1 for further assembly diagnostics.

\pagebreak

\includegraphics[width=\columnwidth]{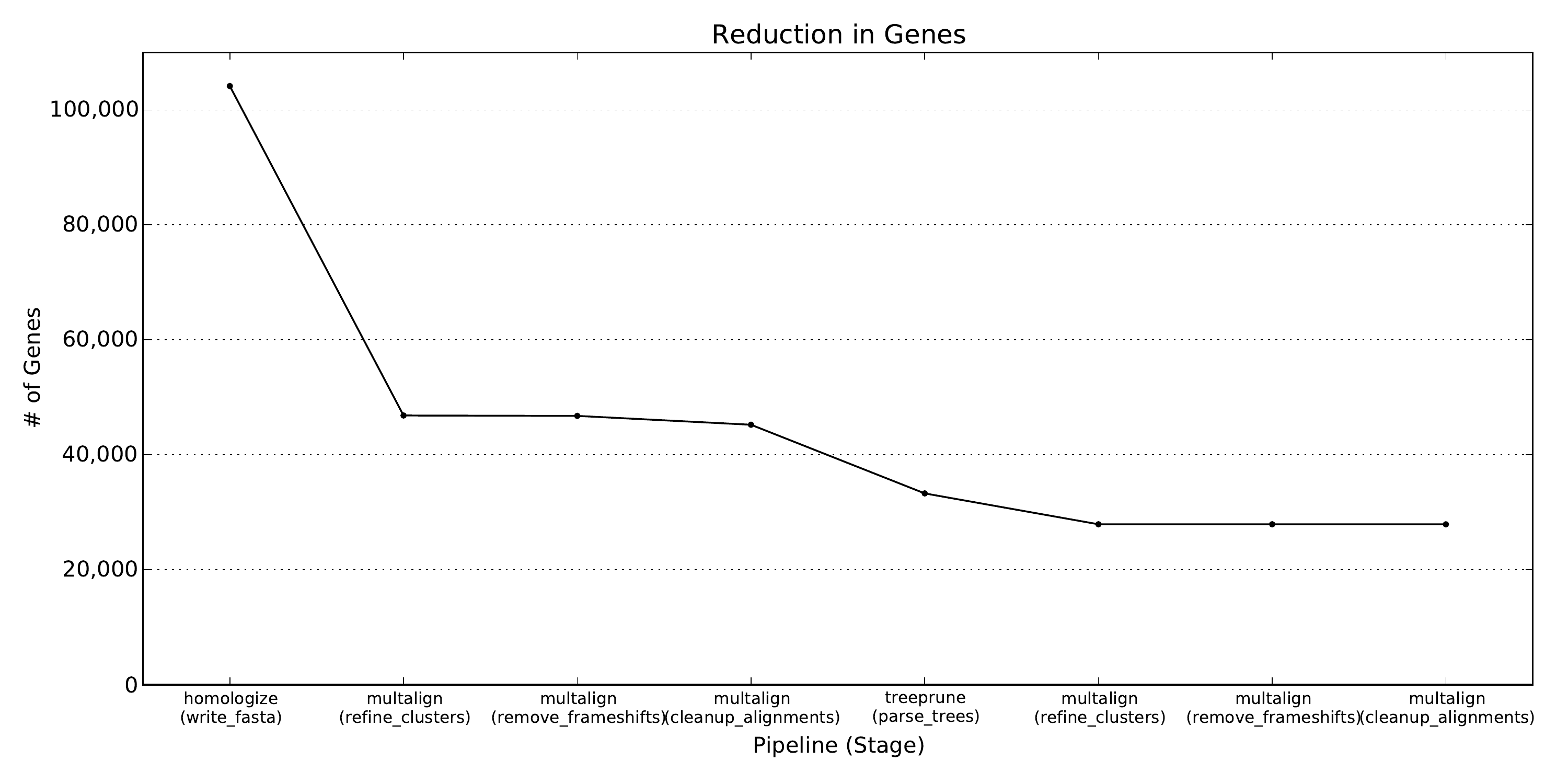}
\subsection*{Figure 2 - The number of gene sequences under consideration at each stage of matrix construction}
Most gene sequences are eliminated in the first step of homology evaluation (the first call to the \texttt{refine\_clusters} stage of the \texttt{multalign} pipeline). Of the remaining sequences, many are eliminated during orthology evaluation (the \texttt{treeprune} pipeline). See Additional File 2 for further diagnostics regarding matrix construction.

\pagebreak

\includegraphics[width=\columnwidth]{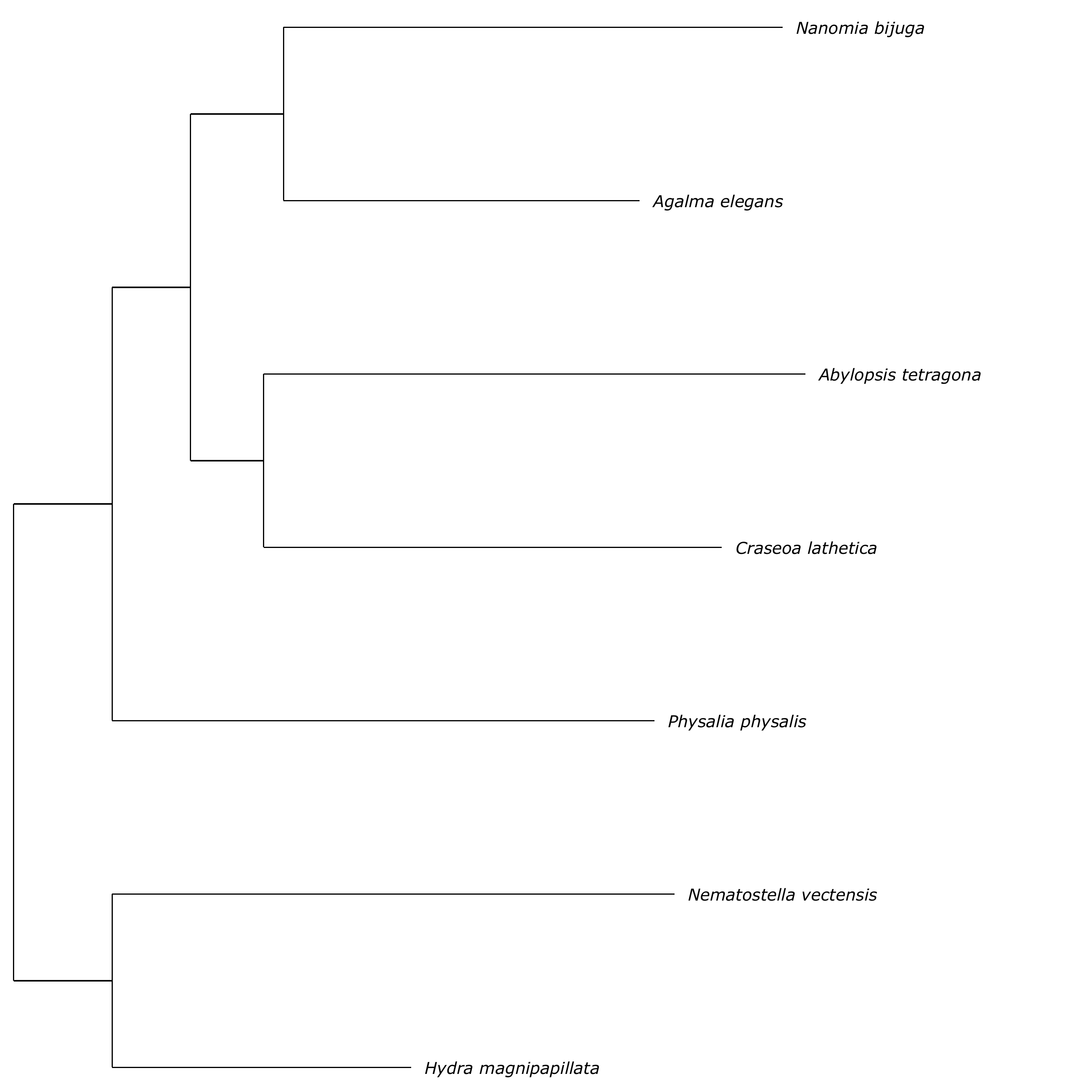}
\subsection*{Figure 3 - The preliminary maximum likelihood phylogeny resulting from the example analysis}
This tree was inferred from the protein supermatrix under the $WAG+\Gamma$ model.

\includegraphics[width=\columnwidth]{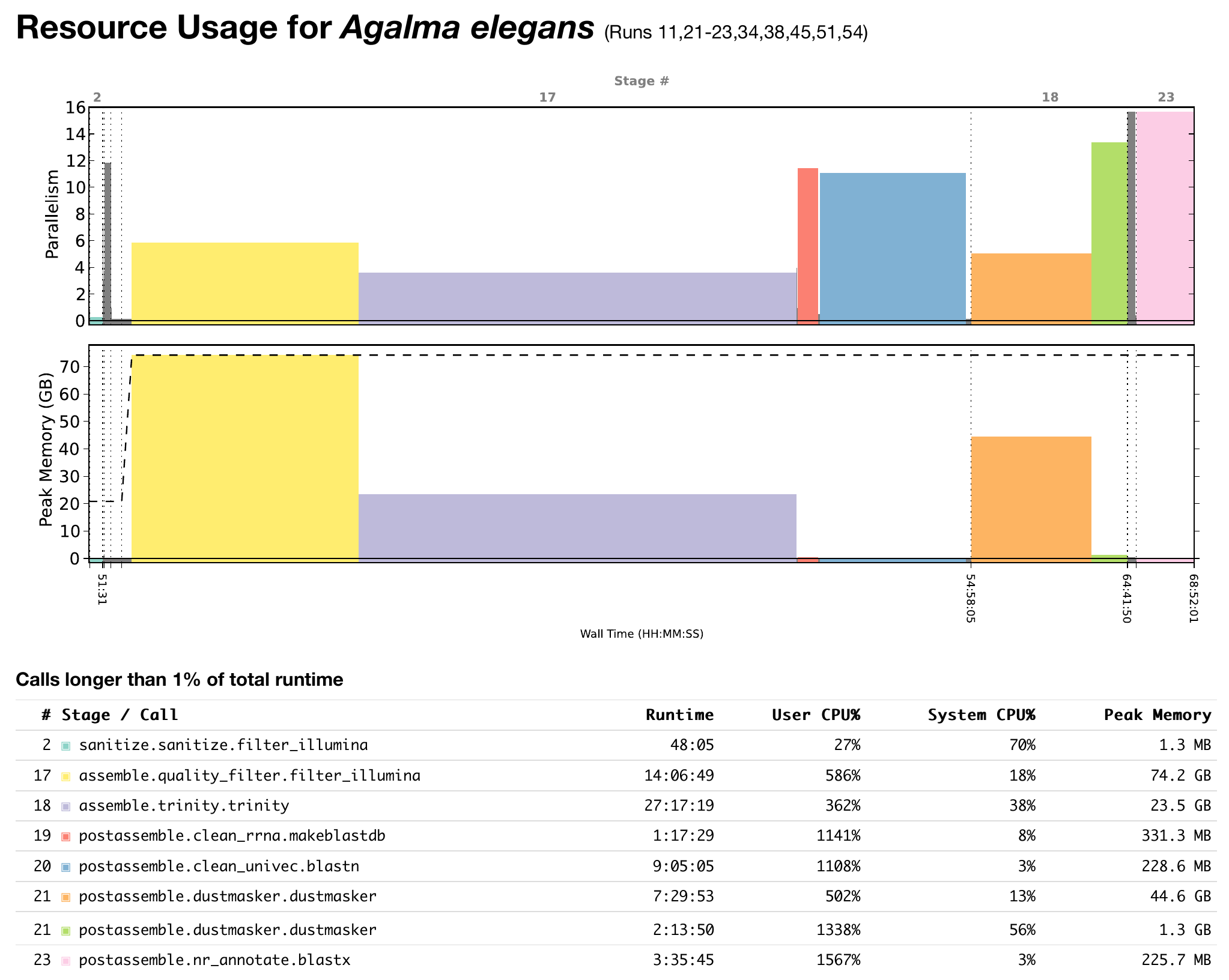}
\subsection*{Figure 4 -  A profile of computational resource utilization for the \texttt{transcriptome} pipeline}
This plot is from the report for the \emph{Agalma elegans} assembly.
\pagebreak




\section*{Additional Files}

  \subsection*{Additional file 1 --- HTML report for assembly of the sample data sets.}
    The HTML report for the assembly of the test data sets from raw reads. The tabular report (index.html) provides an overview across the five assemblies for the ingroup taxa, and includes links (in the Catalog ID column) to detailed reports for the assembly of each species. Fasta files for the annotated transcripts have been removed from the report to reduce file size. This report is available as a zipped archive at \url{https://bitbucket.org/caseywdunn/dunnhowisonzapata2013/downloads/tabular.zip}.

  \subsection*{Additional file 2 --- HTML report for phylogenetic analyses.}
    The HTML report for the phylogenetic analysis of the sample data. This report is available as a zipped archive at \url{https://bitbucket.org/caseywdunn/dunnhowisonzapata2013/downloads/AgalmaExampleTree.zip}.

\end{bmcformat}
\end{document}